\documentclass[12pt, prd, showpacs]{revtex4}
\usepackage{amssymb}
\usepackage{amsmath}

\input{tcilatex}

\begin{document}

\title{Maximum efficiency of the collisional Penrose process}
\author{O. B. Zaslavskii}
\affiliation{Department of Physics and Technology, Kharkov V.N. Karazin National
University, 4 Svoboda Square, Kharkov 61022, Ukraine}
\affiliation{Institute of Mathematics and Mechanics, Kazan Federal University, 18
Kremlyovskaya St., Kazan 420008, Russia}
\email{zaslav@ukr.net }

\begin{abstract}
We consider the collision of two particles that move in the equatorial plane
near a general stationary rotating axially symmetric extremal black hole.
One of the particles is critical (with fine-tuned parameters) and moves in
the outward direction. The second particle (usual, not fine-tuned) comes
from infinity. We examine the efficiency $\eta $ of the collisional Penrose
process. There are two relevant cases here: a particle falling into a black
hole after collision (i) is heavy or (ii) has a finite mass. We show that
the maximum of $\eta $ in case (ii) is less than or equal to that in case
(i). It is argued that for superheavy particles, the bound applies to
nonequatorial motion as well. As an example, we analyze collision in the
Kerr-Newman background. When the bound is the same for processes (i) and
(ii), $\eta =3$ for this metric. For the Kerr black hole, recent results in
the literature are reproduced.
\end{abstract}

\maketitle

\section{Introduction}

Investigation of high-energy collisions of particles near black holes comes
back to Refs. \cite{pir1} - \cite{pir3}. In recent years, interest in this
issue was revived after the observation made by Ba\~{n}ados, Silk and West
(the BSW effect, after the names of its authors) that particle collision
near the Kerr black hole can lead, under certain additional conditions, to
the unbounded growth of the energy in their center of mass $E_{c.m.}$ \cite%
{ban}.\ Later on, in a large series of works, this observation was
generalized and extended to other objects and scenarios. The energy that
appears in the BSW effect is relevant for an observer who is present just
near the point of collision in the vicinity of the black hole horizon.
Meanwhile, what is especially physically important is the Killing energy $E$
of debris after such a collision measured by an observer at infinity. Strong
redshift "eats" a significant part of $E_{c.m.}$, so it was not quite clear
in advance, to what extent the energy $E$ may be high. If $E$ exceeds the
initial energy of particles, we are faced with the energy extraction from a
black hole. This is the so-called collisional Penrose process (the Penrose
process that occurs due to particle collisions).

It turned out that energy extraction from the Kerr black hole is possible
but it was found to be relatively modest \cite{j}, \cite{p}, \cite{cons}. A
more general situation not restricted to the Kerr metric was considered in
Ref. \cite{z} where quite general upper bounds were derived that depend on
the details of the metric. But, also, an indefinitely large $E$ turned out
to be impossible.

The aforementioned results on the energy extraction were obtained for the
standard BSW scenario: both colliding particles move towards a black hole,
one of which is "critical" with fine-tuned parameters (energy and angular
momentum), and the second particle is "usual" (not fine-tuned). However,
numeric findings in Ref. \cite{shnit} showed that if the critical particle
moves away from a black hole, the efficiency of the process significantly
increases and attains 13,92 for the Kerr metric. In what follows, we call
this the Schnittman process (scenario). Later, it was numerically \cite{card}
and analytically \cite{mod} found that if colliding particles move in
opposite directions and both of them are usual, a formally infinite
efficiency becomes possible (the so-called super-Penrose process). The
problem is, however, that a usual particle with a \ finite energy cannot
move away from a black hole although for the critical particle this is
possible (\cite{j}, \cite{z}, \cite{frac}). Therefore, one could think that
an outgoing usual particle could be created in some preceding collision.
However, more careful treatment showed that the kind of particles under
discussion cannot appear as a result of preceding collisions with finite
masses and angular momenta. For a divergent mass of an initial particle this
is possible but this reduces the physical value of the process \cite{pir15}, 
\cite{epl}. One is led to the conclusion that starting from initial
conditions in which usual outgoing particles near the black \ hole horizon
are absent one cannot obtain them by means of additional collisions. As a
result, the super-Penrose process near black holes is impossible. Accounting
for more involved scenarios in which particles that intermediate between the
critical and usual ones participate, only confirmed this conclusion \cite%
{frac}. (There is another option when collision occurs near a white black
hole \cite{gpw} but we do not discuss these rather exotic objects here.)

Quite recently, the numerical estimates of the efficiency of the energy
extraction from the Kerr black hole found in Ref. \cite{shnit} were derived
analytically \cite{ph}, \cite{h}.

In the present work, we consider particle collisions in the Schnittman
scenario near more general stationary rotating axially symmetric black
holes. It turns out that this not only generalizes the aforementioned
results but leads also to some new qualitative possibilities absent for the
Kerr black hole. In particular, the maximum efficiency of the energy
extraction becomes possible even without heavy produced particles that are
necessary in the Kerr case \cite{h}. As an example, we consider collisions
in the Kerr-Newman background. In doing so, colliding particles are assumed
to be neutral. We do not discuss another mechanism of the energy extraction
related to the electrically charged particles \cite{rn}, \cite{esc}.

Throughout the paper, the fundamental constants $G=c=1$.

\section{Basic formulas}

We consider the axially symmetric metric of the form%
\begin{equation}
ds^{2}=-N^{2}dt^{2}+g_{\phi }(d\phi -\omega dt)^{2}+\frac{dr^{2}}{A}%
+g_{\theta }d\theta ^{2}\text{,}
\end{equation}%
where the metric coefficients do not depend on $t$ and $\phi $.
Correspondingly, the energy $E=-mu_{0}$ and angular momentum $L=mu_{\phi }$
are conserved. Here, $m$ is a particle's mass, $u^{\mu }=\frac{dx^{\mu }}{%
d\tau }$ being the four-velocity, $\tau $ the proper time along a
trajectory. We assume that the metric is symmetric with respect to the
equatorial plane $\theta =\frac{\pi }{2}$ and (unless otherwise stated
explicitly) confine ourselves by motion within this plane. Then, without the
loss of generality, we can redefine the radial coordinate in such a way that 
$A=N^{2}$. We do not restrict ourselves by the Kerr or another concrete form
of the metric, so the results are quite general. The equations of motion for
a geodesic particle read%
\begin{equation}
m\dot{t}=\frac{X}{N^{2}}\text{,}
\end{equation}%
\begin{equation}
m\dot{\phi}=\frac{L}{g_{\phi }}+\frac{\omega X}{N^{2}}\text{,}
\end{equation}%
\begin{equation}
m\dot{r}=\sigma Z\text{,}
\end{equation}%
where 
\begin{equation}
X=E-\omega L\text{,}  \label{x}
\end{equation}%
\begin{equation}
Z=\sqrt{X^{2}-N^{2}(m^{2}+\frac{L^{2}}{g_{\phi }})}\text{,}  \label{z}
\end{equation}%
dot denotes differentiation with respect to $\tau $, the factor $\sigma =+1$
or $-1$ depending on the direction of motion.

As usual, we assume the forward-in-time condition $\dot{t}>0$, whence 
\begin{equation}
X\geq 0.  \label{ft}
\end{equation}%
The equality can be achieved on the horizon only where $N=0$.

\subsection{Classification of particles and their properties near the horizon%
}

Particles with $X_{H}>0$ separated from zero are called usual, particles
with $X_{H}=0$ are called critical. If $X_{H}\neq 0$ but is small (of the
order $N$), we call a particle near-critical. Subscript "H" means that the
corresponding quantity is calculated on the horizon. In what follows, we
need the approximate expression for relevant quantities $X$ and $Z$ near the
horizon, where $N\ll 1$. For the near-critical particle,%
\begin{equation}
L=\frac{E}{\omega _{H}}(1+\delta ),  \label{ld}
\end{equation}%
where $\delta \ll 1$. Near the horizon, we assume the Taylor expansion

\begin{equation}
\delta =C_{1}N+C_{2}N^{2}+O(N^{3})\text{,}  \label{d}
\end{equation}%
\begin{equation}
\omega =\omega _{H}-B_{1}N+B_{2}N^{2}+O(N^{3})\text{.}  \label{om}
\end{equation}%
In what follows, we use the notations%
\begin{equation}
b=B_{1}\sqrt{g_{H}}\text{, }h=\omega _{H}\sqrt{g_{H}}\text{, }b_{2}=B_{2}%
\sqrt{g_{H}}.
\end{equation}%
One can find from (\ref{x}), (\ref{z}) and (\ref{ld}) - (\ref{om}) the
expansions for $X$ and $Z$. For the critical particles it reads

\begin{equation}
X=NE(\frac{b}{h}-C_{1})+(C_{1}\frac{b}{h}-\frac{b_{2}}{h}%
-C_{2})EN^{2}+O(N^{3})\text{,}
\end{equation}%
\begin{equation}
Z=Ns+\tau N^{2}+O(N^{3})\text{, }s=s(E,C_{1})=\sqrt{E^{2}[(\frac{b}{h}%
-C_{1})^{2}-\frac{1}{h^{2}}]-m^{2}}\text{,}  \label{s}
\end{equation}%
\begin{equation}
\tau =\frac{E^{2}(\rho +\frac{1}{2h^{2}}\frac{g_{1}}{g_{H}})}{s}\text{,}
\end{equation}%
\begin{equation}
\rho =-C_{1}^{2}\frac{b}{h}+C_{1}C_{2}+C_{1}(\frac{b^{2}-1}{h^{2}}+\frac{%
b_{2}}{h})-C_{2}\frac{b}{h}-\frac{bb_{2}}{h^{2}}\text{.}
\end{equation}%
For the critical one we can put $C_{1}=0$ and obtain%
\begin{equation}
X=NE\frac{b}{h}+O(N^{2})\text{,}
\end{equation}

\begin{equation}
s(E,0)=\sqrt{E^{2}(\frac{b^{2}}{h^{2}}-\frac{1}{h^{2}})-m^{2}}\text{,}
\end{equation}%
\begin{equation}
Z=N\sqrt{E^{2}[\left( \frac{b}{h}\right) ^{2}-\frac{1}{h^{2}}]-m^{2}}%
+N^{2}E^{2}\frac{\frac{1}{2h^{2}}\frac{g_{1}}{g_{H}}-\frac{bb_{2}}{h}}{\sqrt{%
E^{2}(\frac{b^{2}}{h^{2}}-\frac{1}{h^{2}})-m^{2}}}+O(N^{3}).
\end{equation}%
For a usual particle,%
\begin{equation}
X=X_{H}+B_{1}LN-B_{2}LN^{2}+O(N^{3})\text{,}
\end{equation}%
\begin{equation}
Z=X-\frac{N^{2}}{2X}(m^{2}+\frac{L^{2}}{g_{H}})+O(N^{3}).
\end{equation}

\section{General conditions for escaping to infinity}

It is convenient to rewrite (\ref{z}) as

\begin{equation}
Z^{2}=\frac{g_{00}L^{2}}{g_{\phi }}-2\omega LE+E^{2}-N^{2}m^{2}=\frac{g_{00}%
}{g_{\phi }}(L-L_{+})(L-L_{-})\text{,}
\end{equation}%
\begin{equation}
L_{\pm }=\frac{Eg_{\phi }\omega \pm N\sqrt{Y}}{g_{00}},
\end{equation}%
\begin{equation}
Y=(E^{2}+m^{2}g_{00})g_{\phi }\text{.}
\end{equation}

Near the horizon, $g_{00}=-N^{2}+g_{\phi }\omega ^{2}\approx \left( g_{\phi
}\omega ^{2}\right) _{H}>0$.

Then,

\begin{equation}
L_{\pm }=L_{H}+N(L_{H}\frac{b}{h}\pm \frac{\sqrt{\frac{E^{2}}{h^{2}}+m^{2}}}{%
\omega _{H}})+O(N^{2}).  \label{l+-}
\end{equation}

On the horizon,%
\begin{equation}
L_{\pm }=\frac{E}{\omega _{H}}=L_{H}\text{.}
\end{equation}

We are interested in the conditions when a particle can escape to infinity.
In general, they are model-dependent and depend on the behavior of the
metric not only near the horizon but in the intermediate region between the
horizon and infinity as well. However, there are some general necessary
conditions which should be obeyed just near the horizon. In what follows, we
make one more assumption that they are also sufficient (say, there are no
additional maxima of the potential barrier). Here, there are two options.

\begin{equation}
E\geq m,L<L_{H},\sigma =+1,\delta <0,  \label{a}
\end{equation}

\begin{equation}
E\geq m,L_{H}<L<L_{-},\sigma =-1\text{ or }\sigma =+1.  \label{b}
\end{equation}%
In the second case a particle bounces from the turning point before reaching
the horizon. This requires $\delta >0$.

In case (\ref{a}), $C_{1}<0$. In case (\ref{b}), it follows from (\ref{l+-})
that 
\begin{equation}
0<C_{1}\leq C_{m}\text{,}  \label{cm}
\end{equation}%
\begin{equation}
C_{m}=\frac{b}{h}-\frac{\sqrt{\frac{E^{2}}{h^{2}}+m^{2}}}{L_{H}\omega _{H}}=%
\frac{b}{h}-\sqrt{\frac{1}{h^{2}}+\frac{m^{2}}{E^{2}}}.
\end{equation}

\section{Collision near horizon}

Let two particles 1 and 2 collide to produce new particles 3 and 4. The
conservation laws of the energy and angular momentum read

\begin{equation}
E_{1}+E_{2}=E_{3}+E_{4},  \label{e}
\end{equation}%
\begin{equation}
L_{1}+L_{2}=L_{3}+L_{4}.  \label{l}
\end{equation}%
It follows from (\ref{e}) and (\ref{l}) that%
\begin{equation}
X_{1}+X_{2}=X_{3}+X_{4}\text{.}  \label{xx}
\end{equation}%
The conservation of the radial momentum gives us%
\begin{equation}
\sigma _{1}Z_{1}+\sigma _{2}Z_{2}=\sigma _{3}Z_{3}+\sigma _{4}Z_{4}.
\label{zz}
\end{equation}

We concentrate on the following scenario \cite{shnit}. A usual particle 2
comes from infinity, collides with an outgoing particle 1 near the horizon
and produces particles 3 and 4. Particle 4 is usual, it falls down into a
black hole. Particle 3 that escapes to infinity should be near-critical.
This follows from the analysis carried out in \cite{j} for collisions near
the Kerr black hole and in \cite{z} for a much more general case. (It is
also worth noting that an individual particle that moves near the horizon in
the outer region along an outgoing geodesics extendable indefinitely in the
past, should be critical - see, e.g., discussion in Sec. IV A of \cite{frac}%
.) It either goes to infinity immediately after collision or moves inward,
bounces from the potential barrier first and only afterwards escapes to
infinity. Particle 1 is produced from the previous collision and for this
reason should be near-critical as well as particle 3.\ For simplicity, we
assume that particle 1 is exactly critical.\ 

Now, Eq. (\ref{zz}) reads

\begin{equation}
Z_{4}-Z_{2}=\sigma _{3}Z_{3}-Z_{1}\text{.}  \label{z124}
\end{equation}%
Using power expansions for each type of particles and neglecting terms of
the order $N^{2}$ we have

\begin{equation}
\sigma _{3}s(E_{3},C_{1})=A_{1}+E_{3}(C_{1}-\frac{b}{h})\equiv F\text{,}
\label{FA}
\end{equation}%
where $s(E,C)$ is taken from Eq. (\ref{s}),%
\begin{equation}
A_{1}\equiv \frac{b}{h}E_{1}+s(E_{1},0)\text{.}
\end{equation}%
It is easy to find from (\ref{FA}) that%
\begin{equation}
C_{1}=\frac{b}{h}-\frac{A_{1}^{2}+m_{3}^{2}+\frac{E_{3}^{2}}{h^{2}}}{%
2A_{1}E_{3}}\text{.}  \label{y}
\end{equation}%
By substitution back into (\ref{FA}) we obtain%
\begin{equation}
F=\frac{A_{1}^{2}-m_{3}^{2}-\frac{E_{3}^{2}}{h^{2}}}{2A_{1}}\text{.}
\label{F}
\end{equation}

\section{Energy extraction and sign of $\protect\sigma _{3}$}

The energy extraction can be measured by the quantity%
\begin{equation}
\eta =\frac{E_{3}}{E_{1}+E_{2}}\text{.}  \label{ef}
\end{equation}%
We are interested in obtaining the maximum possible value of $\eta $. For
given $E_{1}$ and $E_{2}$, this requires the maximum value of $E_{3}$. This
value obeys bounds that follow from (\ref{FA}) - (\ref{F}). The form of the
bound depends on $\sigma _{3}$.

If $\sigma _{3}=+1$, it follows from (\ref{FA}) and (\ref{F}) that%
\begin{equation}
E_{3}\leq \lambda _{0}\equiv h\sqrt{A_{1}^{2}-m_{3}^{2}}<hA_{1}.
\end{equation}%
If $\sigma _{3}=-1$, $C_{1}\geq 0$ according to (\ref{cm}), and we have from
(\ref{y}) that%
\begin{equation}
E_{3}^{2}-2E_{3}bhA_{1}+h^{2}(A_{1}^{2}+m_{3}^{2})\equiv (E_{3}-\lambda
_{+})(E_{3}-\lambda _{-})\leq 0\text{,}
\end{equation}%
where%
\begin{equation}
\lambda _{\pm }=h(bA_{1}\pm \sqrt{A_{1}^{2}(b^{2}-1)-m_{3}^{2}})\text{,}
\label{la+}
\end{equation}%
\begin{equation}
C_{1}=-\frac{(E_{3}-\lambda _{+})(E_{3}-\lambda _{-})}{2A_{1}E_{3}h^{2}}.
\label{c1}
\end{equation}%
Therefore, 
\begin{equation}
\lambda _{-}\leq E_{3}\leq \lambda _{+}\text{.}
\end{equation}

But $\lambda _{+}>hbA_{1}>hA_{1}$ since the nonnegativity of the square root
in (\ref{la+}) requires $b\geq 1$. Thus $\lambda _{0}<\lambda _{+}$ and the
scenario in which the maximum value of $E_{3}$ is equal to $\lambda _{+}$ is
more effective than that with $E_{3}=\lambda _{0}$. Therefore, in what
follows we concentrate on the case $\sigma _{3}=-1$.

It is implied that after bounce from the potential barrier particle 3
escapes to infinity, so $E_{3}\geq m_{3}$, $\lambda _{+}\geq m_{3}$. It is
clear from (\ref{la+}) that%
\begin{equation}
m_{3}\leq A_{1}\sqrt{b^{2}-1}\text{.}
\end{equation}%
Thus%
\begin{equation}
m_{3}\leq E_{3}\leq \lambda _{+}\text{.}
\end{equation}

When $E_{3}=\lambda _{+}$, inequality $E_{3}\leq \lambda _{+}$ turns into
equality that requires $C_{1}=0$ according to (\ref{c1}).

To make $\lambda _{+}$ as large as possible, we choose $m_{3}=0$. Then,%
\begin{equation}
\left( \lambda _{+}\right) _{\max }=hA_{1}(b+\sqrt{b^{2}-1})\text{.}
\label{max}
\end{equation}%
If also $m_{1}=0$,%
\begin{equation}
A_{1}=E_{1}\frac{(b+\sqrt{b^{2}-1})}{h}\text{,}
\end{equation}%
\begin{equation}
\left( \lambda _{+}\right) _{\max }=E_{1}(b+\sqrt{b^{2}-1})^{2}\text{.}
\label{lam}
\end{equation}

\section{Production of heavy particles}

The subsequent properties of scenarios depend crucially on the value of $%
m_{4}$. This was noticed for the Kerr metric in \cite{h} and is generalized
below. It turns out that if $m_{4}$ is not finite arbitrary quantity but is
adjusted to the location of collision in a special way, one can derive some
universal bounds for the efficiency of the energy extracted. Namely, we
suppose in this Section that $m_{4}^{2}$ has the order $N^{-1}$, so that%
\begin{equation}
m_{4}^{2}=\frac{\mu }{N}+\mu _{0}\text{,}  \label{m4}
\end{equation}%
where $\mu _{0,1}=O(1)$. Then, it follows from (\ref{z}) that

\begin{equation}
Z_{4}\approx \sqrt{X_{4}^{2}-\mu N}\text{,}
\end{equation}%
where we neglected the terms $N^{2}$ inside the square root.

In the above expansion in powers of $N$ that gave rise to (\ref{FA}) it was
tacitly implied that $m_{4}$ was finite. For (\ref{m4}), it should be
somewhat modified. \ \ Eq. (\ref{FA}) is still valid but with another
expression for $A_{1}$:%
\begin{equation}
A_{1}=\frac{b}{h}E_{1}+s(E_{1},0)-\frac{\mu }{2X_{2}}=E_{1}\frac{b+\sqrt{%
b^{2}-1}}{h}-\frac{\mu }{2X_{2}}\text{.}
\end{equation}

Correspondingly, the expression (\ref{max}) also changes. Now,

\begin{equation}
\left( \lambda _{+}\right) _{\max }=E_{1}(b+\sqrt{b^{2}-1})^{2}-\frac{\mu h}{%
2X_{2}}(b+\sqrt{b^{2}-1})<\lambda _{\max }(\mu =0)=E_{1}(b+\sqrt{b^{2}-1}%
)^{2}\text{.}  \label{la}
\end{equation}%
Now, we have

\begin{equation}
\eta \leq \frac{\lambda _{\max }}{E_{1}+E_{2}}\leq \frac{\lbrack b+\sqrt{%
(b^{2}-1)}]^{2}E_{1}}{E_{1}+E_{2}}\leq \eta _{0}\text{,}
\end{equation}%
where%
\begin{equation}
\eta _{0}\equiv \lbrack b+\sqrt{(b^{2}-1)}]^{2}  \label{nmax}
\end{equation}%
is the maximum possible value of $\eta _{0}\,$. Comparison of the terms $%
N^{2}$ in the momentum conservation allows us, in principle, to find $\mu
_{0}$ but we omit this unimportant part.

For the Kerr metric,%
\begin{equation}
b=2\text{, }h=1=b_{2},  \label{kerr}
\end{equation}%
and we obtain $\eta _{0}=(2+\sqrt{3})^{2}\approx 13.92$ that agrees with
previous results \cite{shnit}, \cite{ph}, \cite{h}.

\section{Generalization to nonequatorial motion}

It is interesting that the bound on efficiency found for equatorial motion
admits generalization to the nonequatorial one, provided $m_{4}$ has the
form (\ref{m4}), so particle 4 is heavy. Now, instead of (\ref{z}), we have%
\begin{equation}
Z=\sqrt{X^{2}-N^{2}(m^{2}+\frac{L^{2}}{g_{\phi }}+g_{\theta }\dot{\theta}%
^{2})}\text{.}
\end{equation}

Assuming that $\dot{\theta}^{2}$ are finite for all particles, we see that
only terms of the order $N^{2}$ in the near-horizon expansion can change.\
Meanwhile, the above result was obtained on the basis of the conservation of
the radial momentum (\ref{zz}) and its expansion with respect to $N$ in
which terms of the zero and first order only were taken into account.
Therefore, the bound (\ref{nmax}) remains valid.

\section{Finite $m_{4}$}

\subsection{Restriction on $E_{2}$ from terms of order $N^{2}$}

If the mass $m_{4}$ is finite, the situation becomes much more complex since
one should take into account terms $N^{2}$ in (\ref{zz}). This is because
just these terms can give some lower bound on $E_{2}$ (see below). For the
same reason, generalization to the nonequatorial motion is not
straightforward since it depends on terms $N^{2}$ that are themselves
model-dependent. Let us denoted by $Y_{L}$ and $Y_{R}$ the coefficients at
terms $N^{2}$ in the left and right hand sides of (\ref{z124}) respectively.
Then, direct calculation shows that%
\begin{equation}
Y_{L}=E_{3}(C_{2}-C_{1}\frac{b}{h})+\frac{b_{2}}{h}(E_{3}-E_{1})+\frac{1}{%
2\left( X_{2}\right) _{H}}[m_{2}^{2}-m_{4}^{2}]+\frac{\alpha }{h^{2}}\text{,}
\end{equation}%
where%
\begin{equation}
\alpha =E_{1}-E_{3}+\frac{E_{2}^{2}-(E_{1}-E_{3}+E_{2})^{2}}{2\left(
X_{2}\right) _{H}}.
\end{equation}%
Calculating $Y_{R}$ with arbitrary $m_{3}$ and $\sigma _{3}$, we obtain%
\begin{equation}
Y_{R}=(\frac{g_{1}}{2g_{H}h^{2}}-\frac{bb_{2}}{h^{2}})(-\frac{E_{1}^{2}}{%
\sqrt{E_{1}^{2}\frac{(b^{2}-1)}{h^{2}}-m_{2}^{2}}}+\sigma _{3}\frac{\lambda
_{+}^{2}}{\sqrt{\lambda _{+}^{2}\frac{(b^{2}-1)}{h^{2}}-m_{3}^{2}}})-\sigma
_{3}\frac{\lambda _{+}^{2}C_{2}\frac{b}{h}}{\sqrt{\lambda _{+}^{2}\frac{%
(b^{2}-1)}{h^{2}}-m_{3}^{2}}}.
\end{equation}

We are interested in the case of the potentially maximum efficiency of
extraction, so we put $E_{3}=\lambda _{+}$. It is realized when $C_{1}=0$.
Then, equation $Y_{L}=Y_{R}$ gives us%
\begin{equation}
m_{4}^{2}+2\left( X_{2}\right) _{H}S=m_{2}^{2}+\frac{[E_{2}^{2}-(E_{1}-%
\lambda _{+}+E_{2})^{2}]}{h^{2}}+2\left( X_{2}\right) _{H}(\frac{b_{2}h-1}{%
h^{2}})(\lambda _{+}-E_{1})\text{,}
\end{equation}%
\begin{equation}
S\equiv Y_{R}-E_{3}C_{2}.
\end{equation}%
We find%
\begin{equation}
E_{2}=\frac{1}{2}(\lambda _{+}-E_{1})-\frac{m_{2}^{2}h^{2}}{2(\lambda
_{+}-E_{1})}+Q\text{, }  \label{e2q}
\end{equation}%
where%
\begin{equation}
Q=\frac{h^{2}(m_{4}^{2}+2\left( X_{2}\right) _{H}S)}{2(\lambda _{+}-E_{1})}%
+\left( X_{2}\right) _{H}(1-b_{2}h)\text{.}  \label{Q}
\end{equation}

\subsection{Properties of $S$}

It is convenient to represent the quantity $S$ in the form

\begin{equation}
S=S_{1}+S_{2}\text{,}  \label{S}
\end{equation}%
where%
\begin{equation}
S_{1}=(\frac{bb_{2}}{h^{2}}-\frac{g_{1}}{2g_{H}h^{2}})(-\sigma _{3}\frac{%
\lambda _{+}^{2}}{\sqrt{\lambda _{+}^{2}\frac{(b^{2}-1)}{h^{2}}-m_{3}^{2}}}+%
\frac{E_{1}^{2}}{\sqrt{E_{1}^{2}\frac{(b^{2}-1)}{h^{2}}-m_{2}^{2}}}),
\label{s1}
\end{equation}%
\begin{equation}
S_{2}=-C_{2}\lambda _{+}[\frac{\lambda _{+}\frac{b}{h}\sigma _{3}}{\sqrt{%
\lambda _{+}^{2}\frac{(b^{2}-1)}{h^{2}}-m_{3}^{2}}}+1]  \label{s2}
\end{equation}%
and we put $E_{3}=\lambda _{+}$.

Let us remind that the case of interest (maximum efficiency of collision) is
realized for $\sigma _{3}=-1$. Then, if 
\begin{equation}
\frac{bb_{2}}{h^{2}}-\frac{g_{1}}{2g_{H}h^{2}}>0\text{,}  \label{b2g}
\end{equation}

both $S_{1}>0$, $S_{2}>0$, so 
\begin{equation}
S>0\text{.}
\end{equation}%
We also remind that $\left( X_{2}\right) _{H}>0$ since particle 2 is usual
and the forward-in-time condition (\ref{ft}) should be satisfied. For the
Kerr metric, (\ref{b2g}) is satisfied, $b_{2}h-1=0$ and $Q>0$. However, in
general, (\ref{b2g}) can be violated, $1-b_{2}h$ can have any sign and one
cannot exclude any sign of $Q$ in advance, so both situations should be
considered separately.

\subsection{$Q\geq 0$}

It follows from (\ref{e2q}) that

\begin{equation}
E_{2}\geq \kappa \text{,}
\end{equation}%
where%
\begin{equation}
\kappa =(\frac{y}{2}-\frac{m_{2}^{2}h^{2}}{2y})\text{, }y\equiv \lambda
_{+}-E_{1}\text{.}  \label{ka}
\end{equation}%
As particle 2 comes from infinity, $E_{2}\geq m_{2}$. It makes sense to
consider two subcases separately in the manner close to that in \cite{h}.

At first, let

\begin{equation}
m_{2}\leq \kappa \text{.}  \label{2k}
\end{equation}%
From (\ref{ka}) we have%
\begin{equation}
h^{2}m_{2}^{2}+2m_{2}y-y^{2}\leq 0\text{, }
\end{equation}%
whence%
\begin{equation}
m_{2}\leq m_{+}=\frac{y}{h^{2}}(\sqrt{1+h^{2}}-1)\text{.}  \label{m2}
\end{equation}%
The efficiency of extraction%
\begin{equation}
\eta =\frac{\lambda _{+}}{E_{1}+E_{2}}\leq \frac{\lambda _{+}}{E_{1}+\kappa }%
=\frac{2\lambda _{+}y}{2E_{1}y+y^{2}-m_{2}^{2}h^{2}}=f(m_{2})\text{.}
\label{fm}
\end{equation}%
The function $f$ increases monotonically from $m_{2}=0$ to $m_{2}=m_{+}$.
When $m_{2}=0$,%
\begin{equation}
f(0)=\frac{2\lambda _{+}}{E_{1}+\lambda _{+}}=2\frac{(b+\sqrt{b^{2}-1})^{2}}{%
1+(b+\sqrt{b^{2}-1})^{2}}\text{,}  \label{f0}
\end{equation}%
where (\ref{lam}) was used. For the Kerr metric, $f(0)=\frac{2(2+\sqrt{3}%
)^{2}}{1+(2+\sqrt{3})^{2}}\approx 1.87.$ For $b\gg 1$, $f(0)\approx 2$.

When $m_{2}$ takes a maximum possible value $m_{+}$, by substitution of (\ref%
{m2}) into (\ref{fm}) we find%
\begin{equation}
f(m_{+})=\frac{\lambda _{+}(\sqrt{1+h^{2}}+1)}{E_{1}\sqrt{1+h^{2}}+\lambda
_{+}}\equiv g(\lambda _{+})\text{.}
\end{equation}%
The function $g(\lambda _{+})$ achieves the maximum value at $\lambda
_{+}=\left( \lambda _{+}\right) _{\max }=E_{1}[(b+\sqrt{b^{2}-1})^{2}]$
according to (\ref{lam}). Then,%
\begin{equation}
g[\left( \lambda _{+}\right) _{\max }]\equiv \eta _{1}=\frac{(b+\sqrt{b^{2}-1%
})^{2}(\sqrt{1+h^{2}}+1)}{\sqrt{1+h^{2}}+(b+\sqrt{b^{2}-1})^{2}}<(b+\sqrt{%
b^{2}-1})^{2}=\eta _{0}\text{,}  \label{g}
\end{equation}%
where we took into account that $b\geq 1$, $(b+\sqrt{b^{2}-1})^{2}\geq 1$.

We see that under the condition (\ref{2k}), the efficiency of extraction for
finite masses $m_{4}$ is always less than that for (\ref{m4}).

Now, let 
\begin{equation}
m_{2}>\kappa \text{.}
\end{equation}%
Then,%
\begin{equation}
\kappa <m_{2}\leq E_{2}\text{.}
\end{equation}%
Therefore, we can put $E_{2}=m_{2}$ in (\ref{ef}). For $E_{3}=\lambda _{+}$,
we have%
\begin{equation}
\eta \leq \frac{\lambda _{+}}{m_{2}+E_{1}}\text{.}
\end{equation}%
As this quantity is monotonically decreasing when $m_{2}$ grows, the maximum
is still achieved if $m_{2}=\kappa $, so it coincides with (\ref{g}).

For the Kerr case (\ref{kerr}), we return to the results of \cite{h},%
\begin{equation}
g(\lambda _{+})=\frac{\lambda _{+}}{\left( 2-\sqrt{2}\right) E_{1}+\lambda
_{+}(\sqrt{2}-1)}.
\end{equation}%
When $\lambda _{+}$ takes the maximum posible value $\lambda _{+}=E_{1}(2+%
\sqrt{3})^{2}$, we obtain%
\begin{equation}
\eta _{1}=\frac{(2+\sqrt{3})^{2}}{2-\sqrt{2}+(2+\sqrt{3})^{2}(\sqrt{2}-1)}%
\approx 2.19.
\end{equation}

For dirty black holes with $b\gg 1$, $\eta _{0}\approx 2b$ is also large$.$

\subsection{$Q<0$}

This case has no analog for the Kerr metric. Now,

\begin{equation}
m_{2}\leq E_{2}<\kappa \text{.}
\end{equation}

This requires the validity of (\ref{m2}). Now, we can put $m_{2}=0$ safely
in the expression for $\eta $ and obtain $\eta =\eta _{0}$ according to (\ref%
{nmax}).

It is interesting that now we have not only the upper bound but also the
lower one:%
\begin{equation}
f(m_{2})<\eta <\frac{\lambda _{+}}{E_{1}+m_{2}}\leq \eta _{0}\text{,}
\end{equation}%
where $\eta _{0}$ is given by Eq. (\ref{nmax}). As $f(m_{2})\geq f(0)$, 
\begin{equation}
\eta >2\frac{(b+\sqrt{b^{2}-1})^{2}}{1+(b+\sqrt{b^{2}-1})^{2}}=\frac{2\eta
_{0}}{1+(b+\sqrt{b^{2}-1})^{2}}\equiv \eta _{2}\text{.}
\end{equation}

Thus%
\begin{equation}
\eta _{2}<\eta \leq \eta _{0}\text{.}
\end{equation}

It is seen that always $\eta _{2}>1$, so extraction does occur.

The maximum value $\eta _{0}$ coincides with (\ref{nmax}), so this maximum
is the same for heavy particles and those with finite masses.

\subsection{Massless particles}

The case $Q<0$ implies some relationship between $C_{2}$ and other
parameters. It arises when (\ref{b2g}) is violated or $b_{2}h>1$ (or both).
To avoid cumbersome expressions, let us consider the situation when
particles 2 and 3 are massless or have negligible masses. If one put $%
m_{2}=m_{3}=0$, it is seen from (\ref{s1}), (\ref{s2}) with $\sigma _{3}=-1$
and $\lambda _{+}$ given by (\ref{lam}) that%
\begin{equation}
S_{1}=(bb_{2}-\frac{g_{1}}{2g_{H}})\frac{2E_{1}b(b+\sqrt{b^{2}-1})}{\sqrt{%
b^{2}-1}h}\text{,}
\end{equation}%
\begin{equation}
S_{2}=C_{2}E_{1}\frac{(b+\sqrt{b^{2}-1})}{\sqrt{b^{2}-1}}\text{,}
\end{equation}%
where $C_{2}>0$. Then, according to (\ref{Q}), $Q<0$ entails%
\begin{equation}
0<C_{2}<C_{2}^{(0)}\text{, }  \label{c2}
\end{equation}%
\begin{equation}
C_{2}^{(0)}=-\frac{m_{4}^{2}\sqrt{b^{2}-1}}{2\left( X_{2}\right) _{H}E_{1}(b+%
\sqrt{b^{2}-1})}+D<D\text{,}  \label{cd}
\end{equation}%
\begin{equation}
D=2\frac{T}{h^{2}}\text{, }T=(b^{2}-1)(b_{2}h-1)+bh(\frac{g_{1}}{2g_{H}}%
-bb_{2})\text{.}
\end{equation}

If $D<0,$the coefficient $C_{2}^{(0)}<0$, condition (\ref{c2}) cannot be
obeyed, so $Q\geq 0$.

\section{Kerr-Newman black hole}

Here, we consider an important example of the extremal Kerr-Newman black
hole. Although in astrophysical application the electric charge is quite
small, investigation of the properties of such a black hole has obvious
theoretical interest. For the corresponding metric, one has the following
values of the horizon coefficients relevant in our context ($\theta =\frac{%
\pi }{2}$):

\begin{equation}
\omega _{H}=\frac{a}{M^{2}+a^{2}}\text{, }b=\frac{2a}{M}\text{, }b_{2}=\frac{%
a^{3}}{M^{3}}\text{,}  \label{co}
\end{equation}

\begin{equation}
h=\frac{a}{M},  \label{h}
\end{equation}%
\begin{equation}
\frac{g_{1}}{2g_{H}}-bb_{2}=1-\frac{a^{2}}{M^{2}}-2\frac{a^{4}}{M^{4}},
\end{equation}%
\begin{equation}
b_{2}h-1=\frac{a^{4}}{M^{4}}-1\text{.}
\end{equation}%
These quantities can be obtained by straightforward calculations from the
known metric coefficients. Now, $M^{2}=q^{2}+a^{2}$, where, $q$ is the
electric charge, $a=\frac{J}{M}$, $J$ is the angular momentum.

For simplicity, we assume that $\frac{m_{4}^{2}}{E_{1}\left( X_{2}\right)
_{H}}\ll 1$, so in (\ref{cd}) $C_{2}^{(0)}\approx D$.

It is convenient to introduce a variable $y=\frac{a^{2}}{M^{2}}$. Then,
after some algebra one finds%
\begin{equation}
T=-3y^{2}-2y+1=(y+1)(1-3y)\text{.}
\end{equation}%
Here, the condition $b^{2}\geq 1$ in (\ref{nmax}) requires $y\geq \frac{1}{4}
$, $\frac{a}{M}\geq \frac{1}{2}$ where (\ref{co}) is taken into account.
Also, the existence of a black hole horizon entails $y\leq 1.$ Thus%
\begin{equation}
\frac{1}{4}\leq y\leq 1\text{.}  \label{int}
\end{equation}%
The function $T(y)$ is monotonically decreasing and has one root $y=y_{0}=%
\frac{1}{3}$ that lies within the interval (\ref{int}).

Thus for $\frac{1}{\sqrt{3}}<\frac{a}{M}\leq 1$ the coefficients $T<0$, $D<0$%
. As a result, $C_{2}^{(0)}<0$ and condition (\ref{c2}) cannot be satisfied,
so negative values of $Q$ are forbidden. As $Q\geq 0,$ the energy extraction 
$\eta =\eta _{1}$ is given by (\ref{g}) that now reads

\begin{equation}
\eta _{1}=\frac{\eta _{0}(\sqrt{1+y}+1)}{\sqrt{1+y}+\eta _{0}}\text{, }\eta
_{0}=(2x+\sqrt{4x^{2}-1})^{2}\text{, }x=\frac{a}{M}.  \label{1a}
\end{equation}%
If $\frac{1}{2}\leq \frac{a}{M}\leq \frac{1}{\sqrt{3}}$, $C_{2}^{(0)}\geq 0$%
. Then, if (\ref{c2}) is satisfied, we have $Q<0$ and the energy extraction
is the same for production of massive \ particles and the ones with modest
mass $m_{4},$ so $\eta =\eta _{0}$ according to (\ref{nmax}). If $C_{2}\geq
C_{2}^{(0)}$, the negative value of $Q$ is forbidden again, so extraction is
given by (\ref{1a}).

For slightly positive $C_{2}$, when $y$ increases and passes through the
value $y_{0}$, the maximum efficiency extraction changes abruptly from $\eta
_{0}(y_{0})=3$ to $\eta _{1}(y_{0})=\frac{3(2+\sqrt{3})}{2+3\sqrt{3}}\approx
1.56$.

It is also instructive to compare $\eta _{1}(b,h)$ for the Kerr-Newman and
Kerr metrics. It follows from (\ref{co}) and (\ref{h}) that $b\leq 2$, $%
h\leq 1$, the equality being achieved for the Kerr black hole. We have from (%
\ref{g}) that $\eta _{1}(b,h)\leq \eta _{1}(2,h)$ since $\eta _{1}$ is
monotonically increasing function of $b$ at fixed $h.$ Also, $\eta
_{1}(2,h)\leq \eta _{1}(2,1)$ since $\eta _{1}(2,h)$ is monotonically
increasing function of $h$. Thus $\eta _{1}($Kerr-Newman$)\leq \eta _{1}($%
Kerr$)$.

Remembering also the expression (\ref{nmax}) for $\eta _{0}$, we see that
for the Kerr-Newman black hole the extraction is less effective than for the
Kerr one in both cases (for heavy particles and for particles with finite $%
m_{4}$).

\section{Summary and conclusions}

We considered the Schnittman scenario. In this scenario, an ingoing a usual
particle falling down from infinity collides near the with the outgoing
critical one. It is this scenario in which the efficiency of the energy
extraction from the Kerr black hole was found to increase as compared to the
standard BSW process attaining almost 14 \cite{shnit}. We analytically
derived the upper bound on the extraction efficiency $\eta $ for such a
scenario that is valid not only for the Kerr metric \cite{h}, \cite{ph} but
applies to any rotating stationary axially symmetric black hole. We found $%
\eta _{0}$, the absolute maximum of $\eta $. In this context, one should
distinguish two situations: (i) the mass of a particle that is produced due
to collision and falls into a black hole scales like $m_{4}\sim N^{-1/2}$ in
the point of collision, (ii) $m_{4}=O(1)$. For collisions in the Kerr \
background, $\eta _{0}$ is realized in case (i) only \cite{h} whereas in
case (ii) the allowed maximum value of $\eta =\eta _{1}<\eta _{0}$. However,
we saw that in general the situation is more involved. Depending on the
relation between the parameters of the problem, either the maximum of $\eta $
equal to $\eta _{0}$ is achieved for heavy particles only or $\eta =\eta
_{0} $ can be realized for any $m_{4}$.

The expressions for $\eta _{0}$ and $\eta _{1}$ contain the metric
coefficients and some their first derivatives on the horizon that are
combined in two parameters $b$ and $h$. The results apply to "dirty"
(surrounded by matter) black holes, the Kerr-Newman one, etc. In particular,
we found the intervals of the Kerr-Newman parameter in which both maxima
(for heavy particles and the ones with finite $m_{4}$) can coincide ($\frac{a%
}{M}\leq \frac{1}{\sqrt{3}}$) and those where they cannot ($\frac{1}{\sqrt{3}%
}<\frac{a}{M}\leq 1$). It turned out that the energy extraction for the
Kerr-Newman black hole with $a<1$ is less effective than for the Kerr one ($%
a=1$).

In the case of heavy particles, the bound obtained is valid also for
nonequatorial motion.

Our approach is quite generic in that it is model-independent and can be
used for further investigation of the collisional Penrose process near a
wide class of black holes. It would be of interest to extend it to generic
nonequatorial motion and compare to the approach developed in \cite{ph}.

\begin{acknowledgments}
This work was funded by the subsidy allocated to Kazan Federal University
for the state assignment in the sphere of scientific activities.
\end{acknowledgments}

\end{document}